\documentclass[prd,tightenlines,nofootinbib,showpacs,preprintnumbers,superscriptaddress, twocolumn, longbibliography, notitlepage]{revtex4-1}

\usepackage{amsfonts,amsmath,amssymb,amsthm,bbm,hyperref}
\usepackage{graphicx}
\usepackage{color}

\newcommand{\be}{\begin{equation}}
\newcommand{\ee}{\end{equation}}
\newcommand{\beq}{\begin{eqnarray}}
\newcommand{\eeq}{\end{eqnarray}}

\begin{document}

\title{Quantum creation of a universe-antiuniverse pair}
\author{S. J. Robles-P\'{e}rez}
\affiliation{Canadian Quantum Research Center, 204-3002 32 Ave Vernon, BC V1T 2L7,  Canada}
\affiliation{Estaci\'{o}n Ecol\'{o}gica de Biocosmolog\'{\i}a, Pedro de Alvarado, 14, 06411 Medell\'{\i}n, Spain}

\date{\today}

\begin{abstract}
If one analyses the quantum creation of the universe, it turns out that the most natural way in which the universes can be created is in pairs of universes whose time flow is reversely related. It means that the matter that propagates in one of the universes can be seen, from the point of view of the other universe, as antimatter, and viceversa. They thus form a universe-antiuniverse pair. From a global point of view, i.e. from the point of view of the whole multiverse ensemble, the creation of universes in universe-antiuniverse pairs restores the matter-antimatter asymmetry observed in each individual universe and it might provide us with distinguishable imprints of the whole multiverse proposal.
\end{abstract}

\maketitle



\section{Introduction}\label{sec01}

It has been known for a long time in quantum cosmology that the creation of the universe can be given in pairs. For instance, the Hartle-Hawking no boundary condition \cite{Hartle1983} gives rise a quantum state that can be written in the semiclassical regime as 
\be\label{SPS01}
\phi = \phi^+ + \phi^- \approx e^{+\frac{i}{\hbar} S(a,\varphi)} + e^{-\frac{i}{\hbar} S(a,\varphi)} ,
\ee
where $S(a,\varphi)$ is the Einstein-Hilbert action of a DeSitter like spacetime that is formed from the corresponding Euclidean DeSitter instanton \cite{Hawking1984}. Typically, the components of the superposition state \eqref{SPS01} have been interpreted as representing the contracting and the expanding branches of the DeSitter spacetime. Which component represents the contracting branch and which one represents the expanding one is a matter of convention because, as it is pointed out in Ref. \cite{Rubakov1999}, there is no absolute notion of time in the universe so one can reverse the direction of the time variable and then, $\phi^+$ and $\phi^-$ would interchange their role. However, if one investigates further the appearance of the time variable in the two universes one realises \cite{RP2019c} that the physical time variables of the the two universes represented in \eqref{SPS01} must be reversely related. Then, according to the $CPT$ theorem, the matter content in the two branches must be $CP$ related, where $C$ and $P$ are the charge conjugation and the parity reversal operations, respectively. On the other hand,  it is also shown in Ref. \cite{Rubakov1999} that from the point of view of the thermodynamical arrow of time both branches in \eqref{SPS01} describe an expanding universe. These two reasons make that the superposition state \eqref{SPS01} can more naturally be interpreted as two expanding universes, one of which is filled with matter and the other is filled with antimatter, having these two terms always a relative meaning with respect to each other. The quantum state \eqref{SPS01} can then be interpreted as representing the quantum superposition of a universe-antiuniverse pair \cite{RP2019c}.

The creation of the universe in a universe-antiuniverse pair would thus restore the matter-antimatter asymmetry observed from the point of view of the single universe. The idea of a time reversal relation between a pair of universes to explain the matter-antimatter asymmetry observed in our universe is not new, actually. It dates back at least to the early 70's \cite{Albrow1973} and it was even posed by Sakharov in the early 80's too \cite{Sakharov1980}. However, for some reason these models have not received the attention they deserve. One of these reasons may be that the consideration of other universes has typically been considered an unphysical or a metaphysical proposal in the sense of being unobservable and therefore untestable. The idea behind the rejection can be sketched as follows: on the one hand, if some event is observable, then, there is a time-like or null path joining together the original event and the observation event and, thus, these two events belong to the same universe; and, on the other hand, if a given event belongs to a different universe, which from the classical point of view is disconnected to the observer's universe, then, any two events of the two universes cannot be joined by a time like or null path so the original event cannot be observed. Thus, the multiverse has typically been considered as a non falsifiable proposal.

However, at least three caveats must be raised at this point. First, a theoretical consistency of the theory is an important sign to at least taking the proposal into consideration. After all we can infer the existence of an otherwise unobservable stellar object (say a black hole) from the theoretical consistency of the perturbed motion of the observable companion, and symmetry consistencies made theoretical physicists to predict the existence of the charm quark; not to talk about the unobserved 'dark matter' that is basically supported by consistency arguments. Second, observability and falsifiability are not the same thing, as it is clearly argued\footnote{Tegmark poses the following example: \emph{a theory stating that there are 666 parallel universes, all of which are devoid of oxygen, makes the testable prediction that we should observe no oxygen here, and is therefore ruled out by observation}, cfr. Ref. \cite{Tegmark2007}, p. 105.} in Ref. \cite{Tegmark2007} (see also, Ref. \cite{Alonso2019} for a recent review). Third, the classical argument exposed above rejecting the multiverse is the typical classical way of thinking that has constantly been challenged by the quantum theory from the very beginning (let us note, for instance, the well-known EPR paradox \cite{Einstein1935}). More concretely, the direct non-observability, in the classical sense, does not exclude the possibility of measuring observable effects derived from the existence of quantum correlations or entanglement between the state of some matter field in two distant places. For instance, in Ref. \cite{RP2018b} it is shown with the help of the parametric amplifier setup of quantum optics that an isolated observer can infer the existence of an unobservable partner mode of the radiation field only from the photon number distribution of the light beam that the observer detects. Similarly, one can show \cite{RP2018a} that the existence of a partner antiuniverse would leave not only observable but also distinguishable imprints in the properties of a universe like ours, making falsifiable the creation of universes in universe-antiuniverse pairs as well as the whole multiverse proposal.

This paper is outlined as follows. In Sect. \ref{sec02} we shall review the paradigmatic example of the creation of a DeSitter spacetime. We shall obtain the quantum state \eqref{SPS01} and interpret it as the superposition state that represents an expanding and a contracting universes, as usual. In Sect. \ref{sec03} we shall analyse the matter content of the universes and the appearance of the \emph{physical} time variable, i.e. the one that appears in the Schr\"odinger equation. We shall show that the physical time variables of the two universes must be reversely related and that, in terms of the time variable measured by the inhabitants of the universe, both universes are expanding universes with the observer's universe initially filled of matter and the partner universe initially filled with antimatter. In Sect. \ref{sec04} we shall review the kind of observable imprints that the creation of the universe in a universe-antiuniverse pair should leave, and finally, in Sect. \ref{sec05} we shall briefly draw some conclusions.


\section{Quantum creation of the universe}\label{sec02}

The dynamics of the gravitational field can be obtained in the Lagrangian framework from the variational principle of the Einstein-Hilbert action, which in the canonical form is essentially the integral over the spacetime manifold $\mathcal M$ of the Ricci scalar plus some boundary term (for the details see, for instance, Ref.  \cite{Kiefer2007}). The essence of Einstein's geometrodynamics is the foliation of the spacetime by a set of spatial sections distributed along the time variable. In that case, the evolution of the universe can be seen as the time evolution of the metric tensor $h_{ij}(t)$ of the $3$-dimensional spatial hypersurface\footnote{As Wheeler says \cite{Wheeler1968}, \emph{Eintein's geometrodynamics deals with the dynamics of $3$-geometry, not $4$-geometry!} (emphasis his).} $\Sigma_t$. To this action one must add the action of the matter fields that propagate in the background spacetime. They together form the total action from which one can obtain the field equations of the whole universe. In general, they are very complicated if not impossible to solve. However, in cosmology we are mainly interested in describing a universe that is created with some degree of symmetry. We know that the fluctuations of the gravitational field are length dependent and they become of order of the metric tensor at the Planck length \cite{Wheeler1957}. Therefore, if we assume that the universe is created with a length scale well above from the Planck length, then, it can be described, at least as a first approximation, by a homogeneous and isotropic metric with small inhomogeneities propagating therein.

Therefore, let us consider the FRW metric
\be
ds^2 = - N^2 dt^2 + a^2(t) d\Omega_3^2 ,
\ee
where $a(t)$ is the scale factor and $d\Omega_3^2$ is the line element on the $3$-sphere of unit radius\footnote{We are assuming closed spatial sections of the spacetime.}; and a scalar field representing the matter content of the universe given by,
\be
\varphi(t,\vec x) = \varphi(t) + \sum_{\textbf n} f_{\textbf n}(t) Q^{\textbf n}(\vec x) ,
\ee
where $\varphi(t)$ is the homogeneous mode, $Q^n(\vec x)$ is the scalar harmonic on the $3$-sphere, and  $f_{\textbf n}(t)$ represent the inhomogeneities of the matter field. The homogeneous mode contains the major part of the energy of the matter field and contributes to the evolution of the background spacetime, and the inhomogeneities can be seen, at least for the modes with a large value of $n \equiv |{\textbf n}|$, as the particles of the field that propagate in an evolving background spacetime. If the inhomogeneities are sufficiently small the total action decouples and can be written as \cite{Kiefer1987, RP2018a}
\beq\label{ACT01}
S=\frac{1}{2} \int dt N \left( - \frac{a \dot a^2}{N^2} + a - H^2 a^3 \right) \\ + \frac{1}{2} \int dt N a^3 \sum_{\textbf n} \frac{\dot f_{\textbf n}^2}{N^2} - \omega_n^2 f_{\textbf n}^2 ,
\eeq
where
\be\label{OME01}
\omega_n^2=\frac{n^2-1}{a^2} + m^2  ,
\ee
with $m$ the mass of the scalar field. The first term in \eqref{ACT01} is the action of the background spacetime. The time derivative of the homogeneous mode of the scalar field does not appear because we have assumed the typical conditions for the initial inflationary stage of the universe, $\dot\varphi \ll 1$, and, $ 2 V(\varphi_0) \equiv H^2 \gg 1$, in Planck units. The second term in \eqref{ACT01} is the action of a set of uncoupled harmonic oscillators with time dependent 'mass', given by $M=a^3(t)$, and time dependent frequency given by \eqref{OME01}. The lapse function must be retained in \eqref{ACT01} until  the variation of the action with respect to $N$ is performed, and then will be set to one.

Let us first consider the dynamics of the background spacetime by neglecting the inhomogeneities. In that case, the invariance of the action with respect to the lapse function gives rise the classical Hamiltonian constraint, 
\be\label{H001}
\mathcal H_0 = \frac{1}{2a} \left( -p_a^2 + H^2a^4 - a^2 \right) = 0 ,
\ee
which in terms of the time derivative of the scale factor, $p_a = -a \dot a$, can be written as
\be\label{SF01}
\dot a = \sqrt{H^2 a^2 - 1} .
\ee
It has  the well-known solution, $a(t) = a_0 \cosh H t$, that represents a universe that contracts from an infinite volume until it reaches the minimum volume element, given by $a_0^3$, and then starts expanding again. For this reason, the two branches of the solution \eqref{SF01} are called the \emph{contracting} and the \emph{expanding} branches of the universe (see, Fig. \ref{figure01}).

Quantum mechanically, the quantum state of the universe is represented by the wave function that is the solution of the Wheeler-DeWitt equation obtained from the canonical quantisation of the classical momentum in \eqref{H001}, $p_a \rightarrow - i\hbar \frac{\partial}{\partial a}$, i.e.
\be\label{WDW01}
\hbar^2 \frac{\partial^2\phi(a)}{\partial a^2} + \Omega^2(a) \phi(a) =  0 ,
\ee
where
\be
 \Omega^2(a) = H^2a^4 - a^2 .
\ee
We do not know the exact solutions of \eqref{WDW01} but far from the turning point, $a_0 = H^{-1}$, we can use the WKB wave functions. Moreover, the turning point splits the minisuperspace in two parts with two different regimes for the wave function $\phi(a)$. For the value $a > a_0$, the wave function is in the oscillatory regime with WKB solutions given by the complex exponentials, $\phi^\pm \propto e^{\pm \frac{i}{\hbar} S(a)}$, where $S(a) = \int \Omega(a') da'$. On the other hand, the value, $a < a_0$, defines the tunnelling region of the minisuperspace where the wave function is given by a linear combination of the real exponentials\footnote{The universe is said then to be created 'from nothing' meaning by that that it is created from a quantum tunnelling process into the classically allowed region of the minisuperspace.}, $e^{\pm \frac{1}{\hbar} I(a)}$, with $I(a) = \int |\Omega(a')| da'$. The exact combination of wave functions depends on the boundary condition imposed on the state of the universe. For instance, with the Hartle-Hawking no boundary proposal \cite{Hartle1983}, the quantum state of the universe in the oscillatory region turns out to be
\be\label{HH01}
\phi(a)=\phi^+ + \phi^- \approx \frac{1}{\sqrt{\Omega}} e^{+\frac{i}{\hbar} S(a)} + \frac{1}{\sqrt{\Omega}} e^{-\frac{i}{\hbar} S(a)} .
\ee
The customary interpretation of the wave function \eqref{HH01} is that it represents two universes, which according to the relation
\be\label{CM01}
-a\dot a= p_a \approx \langle \phi^\pm | \hat p_a |\phi^\pm \rangle \sim \pm\frac{\partial S}{\partial a} \ \Rightarrow \ \dot a = \mp \frac{1}{a} \frac{\partial S}{\partial a} ,
\ee
one, $\phi^-$,  is expanding and the other, $\phi^+$, is contracting. A decoherence process makes that the two universes can rapidly be considered independently \cite{Halliwell1989, Kiefer1992}. The typical choice is then to consider the expanding branch as the representative of our universe and disregard the contracting one as not being physically significant. However, we shall see in the next sections that the two branches may form a non-separable state with important consequences.


\section{Matter-antimatter content of the universe}\label{sec03}

\begin{figure}
\centering
\includegraphics[width=7 cm]{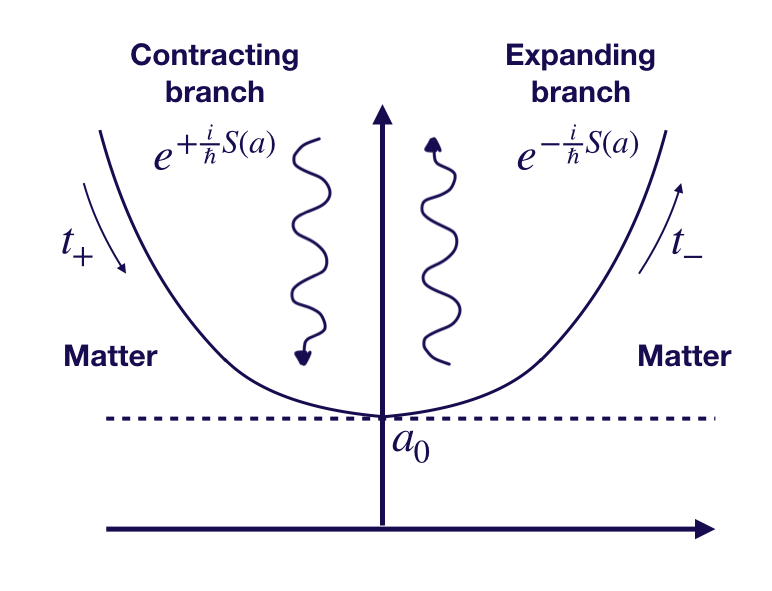}
\caption{The contracting and the expanding branches of a DeSitter spacetime, both filled with matter.}
\label{figure01}
\end{figure}

Let us now analyse the matter content of the universe by considering the total Hamiltonian constraint,
\be\label{H101}
\left(\hat{\mathcal H}_0 + \hat{\mathcal H}_m \right) \phi(a,  f_{\textbf n})=0 ,
\ee
where $\hat{\mathcal H}_m$ is the Hamiltonian of the inhomogeneities of the matter field. The solution of \eqref{H101} is not much different to the wave function \eqref{HH01}. It contains now a factor that gathers all the dependence on the inhomogeneous degrees of freedom,
\be\label{HH02}
\phi^\pm(a, f_{\textbf n})=  \frac{1}{\sqrt{\Omega(a)}} e^{\pm\frac{i}{\hbar} S(a)} \psi_\pm(a, f_\textbf{n}) .
\ee
It comes now one of the most beautiful features of quantum cosmology, the appearance of the classical time variable and the quantum dynamics of the matter fields. If one inserts the wave function \eqref{HH02} into the Wheeler-DeWitt equation \eqref{H101}, assumes that the background spacetime is varying very slow compared with the variation of the matter fields, and solves it order by order in $\hbar$, one obtains at order $\hbar^0$ the Hamilton-Jacobi equation
\be\label{HJ01}
-\left( \frac{\partial S}{\partial a} \right)^2 + \Omega^2 = 0 ,
\ee
which is equivalent to the classical momentum constraint \eqref{H001} if one identifies the momentum conjugated to the scale factor $p_a$ with $\pm \frac{\partial S}{\partial a}$. In that case, Eq. \eqref{HJ01} shows that the solutions for the classical momentum of the background spacetime are, $p_a = \pm \Omega$. These values of the momentum $p_a$ are associated to the two branches in \eqref{HH01} so the creation of universes in pairs would thus conserve the total amount of momentum conjugated to the scale factor. Furthermore, we can define a time variable, $t_\pm$, called the WKB time \cite{Kiefer1987}, as
\be\label{WKBt01}
\frac{\partial }{\partial t_\pm} = \mp \frac{1}{a} \frac{\partial S}{\partial a} \frac{\partial }{\partial a}  ,
\ee
in terms of which one recovers the classical Friedmann equation \eqref{SF01},
\be\label{FE02}
\dot a = \mp \frac{1}{a} \frac{\partial S}{\partial a} = \mp \sqrt{H^2 a^2 - 1} .
\ee 
On the other hand, at order $\hbar^1$ in $\mathcal H_0$ one obtains
\be\label{SCH00}
\mp i \hbar \frac{1}{a} \frac{\partial S}{\partial a}\frac{\partial }{\partial a} \psi_\pm = \hat{\mathcal H}_m \psi_\pm ,
\ee
which is essentially a Schr\"odinger like equation if one realises that the l.h.s. is basically the derivative of the wave function $\psi_\pm$ with respect to the time variable of the classical background defined in \eqref{WKBt01}. However, there is a freedom in the choice of the sign in \eqref{WKBt01} that has to be analysed carefully. As we have said, in terms of the cosmic time $t$, the wave function $\phi^+$ represents a contracting universe and $\phi^-$ an expanding universe [see, \eqref{CM01}]. In that case, in order for the WKB-time \eqref{WKBt01} to represent the cosmic time $t$, we have to choose the variable $t_-$ in the branch $\phi^-$, so that 
\be\label{SF41-}
\frac{\partial a}{\partial t_-}  = \frac{1}{a} \frac{\partial S}{\partial a} > 0 ,
\ee
represents an expanding universe; and for the contracting branch represented by $\phi^+$ we must choose $t_+$, so that
\be\label{SF41+}
\frac{\partial a}{\partial t_+}  = - \frac{1}{a} \frac{\partial S}{\partial a} < 0,
\ee
describes a contracting universe. With this choice, the equation \eqref{SCH00} reads,
\be\label{SCH052a}
i \hbar \frac{\partial }{\partial t_\pm} \psi_\pm(t_\pm, \varphi) = \hat{\mathcal H}_m \psi_\pm(t_\pm, \varphi) ,
\ee
where, $\psi_\pm(t_\pm, \varphi) \equiv \psi_\pm[a(t_\pm), \varphi]$, evaluated in the solutions of the background given by \eqref{SF41-} and \eqref{SF41+}. Therefore, we have ended up with two universes, one contracting and another expanding, both filled with matter, which is the customary interpretation (see, Fig. \ref{figure01}).

\begin{figure}
\centering
\includegraphics[width=7 cm]{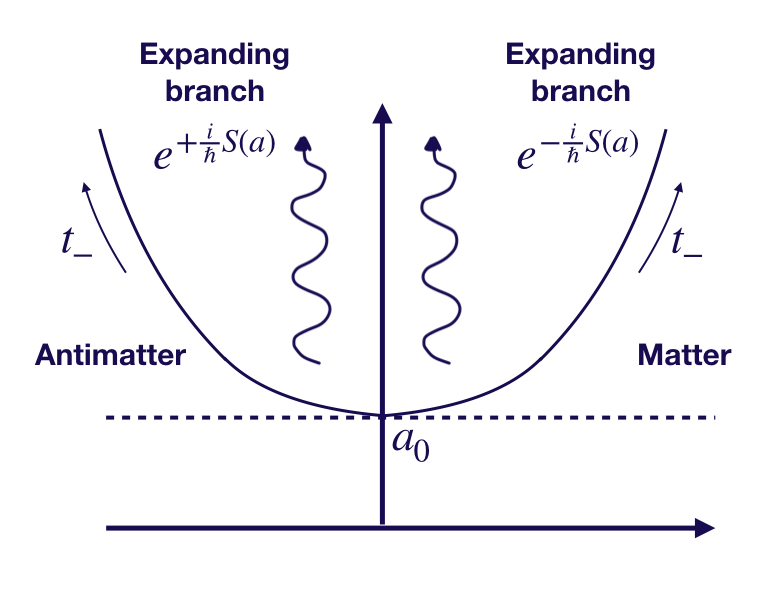}
\caption{In terms of the internal time $t_-$ the two branches can be interpreted as expanding universes, one of them filled with matter and the other filled with antimatter.}
\label{figure02}
\end{figure}

We can make however a different interpretation. It can be assumed that the physical time variable, i.e. the time variable measured by actual clocks that are eventually made of matter, is the time variable that appears in the Schr\"odinger equation of the observer's physical experiments. In that case, it is worth noticing that the physical time variable of observers in the two universes is reversely related, $t_+ = - t_-$. For instance, let us consider $t_-$ as the physical time\footnote{We are then assuming the time measured by one particular observer.}. Then, in terms of the time variable $t_-$ the evolution of the scale factor is given by \eqref{SF41-} so the two wave functions, $\phi^+$ and $\phi^-$, represent both an expanding universe. However, in terms of $t_-$ the Schr\"odinger equation \eqref{SCH00} becomes,
\be\label{SCH052a1}
i \hbar \frac{\partial }{\partial t_-} \psi_-(t_-, \varphi) = \hat{\mathcal H}_m \psi_-(t_-, \varphi) ,
\ee
for the wave function $\psi_-$, and
\be\label{SCH052a2}
- i \hbar \frac{\partial }{\partial t_-} \psi_+(t_-, \varphi) = \hat{\mathcal H}_m \psi_+(t_-, \varphi) ,
\ee
for the wave function $\psi_+$. The 'wrong sign' in \eqref{SCH052a2} is not problematic \cite{Rubakov1999}. It only indicates that \eqref{SCH052a2} is the Schr\"odinger equation of the complex conjugated wave function $\psi_+^*$ with a $CP$-transformed Hamiltonian \cite{Rubakov1999}. It is therefore the Schr\"odinger equation of the conjugated field that represents the antimatter of $\varphi$. In this case, we have ended up with the description of two expanding universes, one of them filled with matter and the other filled with antimatter (see, Fig. \ref{figure02}).

\begin{figure}
\centering
\includegraphics[width=7 cm]{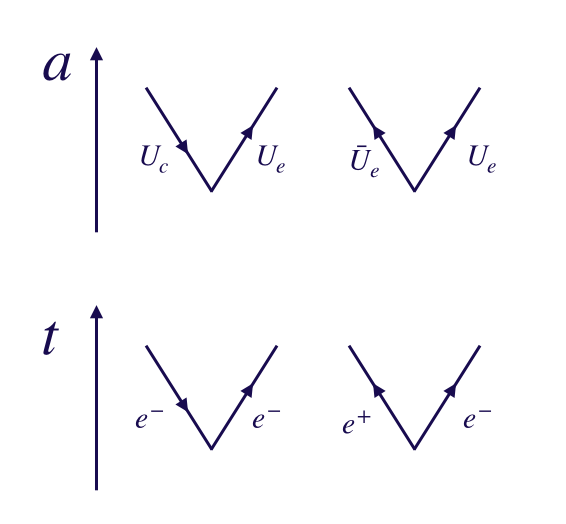}
\caption{In analogy to the creation of particle-antiparticle pairs in a quantum field theory, a contracting and an expanding pair of universes can be interpreted as an expanding universe-antiuniverse pair in the third quantisation formalism.}
\label{figure03}
\end{figure}

The two interpretations can be graphically sketched as in Fig. \ref{figure03}. It clearly resembles the interpretation of particles and antiparticles in a quantum field theory of matter fields (e.g. QED). The analogy can be taken further and the creation of the universe can be more appropriately described in the field theoretical approach called \emph{third quantisation} \cite{DeWitt1967, McGuigan1988, Rubakov1988, RP2012a}, where the wave function of the universe can be seen as a field that propagates in the superspace, in which the time like variable is the volume of the universes\footnote{The full theoretical description of the wave function $\hat \phi$ will be published soon.}. Therefore, the positive and negative frequency modes (the 'particles' and 'antiparticles') can be associated with expanding and contracting universes, or following an interpretation more consistent with the field theoretical approach they can be interpreted as expanding universe-antiuniverse pairs (see, Fig. \ref{figure02}).


\section{Observational imprints}\label{sec04}

One of the most interesting properties of the creation of the universe in universe-antiuniverse pairs is that besides restoring the matter-antimatter asymmetry apparently perceived from the point of one of the single universes, it might provide us as well with observational imprints in the properties of a universe like ours originated in the entanglement of the matter and antimatter fields of the two universes. The quantum field theory of the matter field in the two universes would follow the customary approach and can be expanded in Fourier modes as usual,
\be
\varphi(x,t) =\frac{1}{\sqrt{2}}\sum_\textbf{n} F_\textbf{n}(x) \, v^*_n(t) \, \hat a_\textbf{n} + F^*_\textbf{n}(x) \, v_n(t) \, \hat b^\dag_{-\textbf{n}} .
\ee
The only difference with respect to the development in a single universe is that now the particles ($\hat a_\textbf{n}$) and the antiparticles ($\hat b_\textbf{n}$) would live (propagate) in different but correlated universes. In a time evolving spacetime there is a generation of particles along the evolution of the universe because the invariant representations, $\hat a_\textbf{n}$ and $\hat b_\textbf{n}$, do not coincide with the diagonal representation of the Hamiltonian at any given time. Both representations, the invariant and the instantaneously diagonal representation, are related by a Bogolyubov transformation. However, in the case of a universe-antiuniverse pair, due to the common origin, one can assume\footnote{This can also be seen as a plausible boundary condition.} that the modes of the two universes are entangled so the Bogolyubov transformation would then read \cite{RP2018a}
\beq
\hat a_\textbf{n} &=& \mu(t) \, \hat c_\textbf{n} - \nu^*(t) \, \hat d^\dag_{-\textbf{n}}  \\
\hat b_{\textbf{n}} &=& \mu(t) \, \hat d_{\textbf{n}} - \nu^*(t) \, \hat c^\dag_{-\textbf{n}}  ,
\eeq
where $\mu(t)$ and $\nu(t)$ are two functions that for simplicity we omit here (see, Ref. \cite{RP2018a} for the details). In that case, the composite vacuum state of the invariant representation, $|0_a0_b\rangle$, would be full of particles and antiparticles that would live in disconnected universes so they would not annihilate each other.

The quantum state of the matter field in one of the two universes would be given by the density matrix that is obtained by tracing out from the composite vacuum state, $\rho = |0_a0_b\rangle \langle 0_a 0_b |$, the degrees of freedom of the partner universe, i.e. \cite{RP2018a}
\be\label{RD01}
\rho_1 = {\rm Tr}_2\rho =  \prod_\textbf{n} \frac{1}{Z_\textbf{n}} \sum_{N=0}^\infty e^{-\frac{1}{T} (N +1/2)} | N_{c,\textbf{n}} \rangle \langle N_{c,\textbf{n}} | ,
\ee
where $| N_{c,\textbf{n}} \rangle$ are the number states of the diagonal representation in one of the universes, $Z_\textbf{n}$ is the partition function, and
\be
T \equiv T_n(t) = \frac{1}{\ln(1+\frac{1}{|\nu|^2})} .
\ee
The quantum state \eqref{RD01} is a very specific state so in principle one should expect some distinguishable imprints from it. Let us notice that it is not exactly a thermal state for two reasons. First, the temperature of entanglement $T$ is time dependent\footnote{This is however not an important departure from the thermal state in the sense that at any given time the number of particles of the matter field follows a thermal distribution}; and second and more important, it depends on the value of the mode, i.e. the modes have not thermalised in \eqref{RD01} so we cannot properly talk about a thermal state. In fact, it can be shown \cite{RP2018a} that $T_n \rightarrow 0$ for large modes ($n \gg 1$) meaning that the local particles of the field do not feel the inter-universal entanglement; only the modes with wavelengths of order of the Hubble distance are significantly affected. The quasi thermal character of the quantum state \eqref{RD01} is a very specific prediction of the creation of the universe in a universe-antiuniverse pair.

With the quantum state \eqref{RD01} one can compute all the associated thermodynamical magnitudes. For instance, the energy of the state \eqref{RD01} is
\be
E_1 = {\rm Tr} \hat \rho_1 \hat{\mathcal H}_1 = \frac{\omega_n}{2} \left( |\mu_n|^2 + |\nu_n|^2 \right) ,
\ee
which in the case of a flat DeSitter spacetime produces a backreaction energy density given by \cite{RP2018a}
\be
\varepsilon =  \frac{H^4}{8} \left\{ 1 - \frac{m^2}{H^2} \log\frac{b^2}{H^2} + \left( 1+\frac{m^2}{H^2} \right) \left( 1-\frac{b^2}{H^2} \right) \right\} ,
\ee
where $b$ is an infrared cutoff. However, it turns out that this energy is the same that the one produced by the backreaction of the superhorizon modes of the field in the single universe scenario (see \cite{Mersini2008c, Mersini2008d}). Therefore, it is an observable imprint of the creation of universes in pairs but it is not a distinguishable one.

A distinguishable imprint may come from the spectrum of fluctuations of the matter field. In the customary scenario of a single universe it is typically given by \cite{Mukhanov2007}
\be
\delta\phi_\textbf{n} = \frac{H}{\sqrt{8\pi}} x^\frac{3}{2} \left( \mathcal{J}_q^2(x) + \mathcal{Y}_q^2(x) \right)^\frac{1}{2} ,
\ee
where,
\be
x \equiv \frac{n }{H a} = \frac{n_\text{ph}}{H} \sim \frac{H^{-1}}{L_\text{ph}} .
\ee
However, if the initial state of the inhomogeneities is given by the quasi-thermal state \eqref{RD01}, then \cite{RP2018a}
 \be
 \langle |\phi_n|^2 \rangle = \frac{1}{M \omega_n}  (|\nu|^2 + \frac{1}{2})   ,
 \ee
 and the spectrum of fluctuations, which is given by
\be\label{FLUCT00}
\delta\phi_\textbf{n} = \frac{n^\frac{3}{2}}{2 \pi} \Delta\phi_n ,
\ee
with
\be
(\Delta\phi_n)^2 = \langle |\phi_n|^2 \rangle - |\langle \phi_n \rangle|^2 .
\ee
can be related to the spectrum of fluctuations in the single universe scenario by \cite{RP2018a}
 \be\label{QF01}
 \frac{\delta\phi_\textbf{n}^{eu}}{\delta\phi_\textbf{n}^{su}} = \sqrt{\frac{1}{2}\left( 1 + \frac{x^2}{(1+x^2 )(1+\frac{m^2}{H^2 x^2})} \right) }  ,
\ee
where the superscripts '$eu$' and '$su$' refer to the  entangled universe and the single universe scenarios, respectively. Let us first notice that the  large modes ($x\gg 1$) are in the vacuum state and then, $\delta\phi_\textbf{n}^{th} \approx \delta\phi_\textbf{n}^{I}$, as expected (large modes do not feel the inter-universal entanglement). However, the departure may be significant for the horizon modes, $x \sim 1$. This is a distinctive effect of the creation of the universes in entangled universe-antiuniverse pairs and it should leave, at least in principle, an observable imprint in the properties of the CMB.  It has no analogue in the context of an isolated universe and therefore it is a distinguishable effect of the creation of universes in pairs that, incidentally, would make falsifiable the whole multiverse proposal.


\section{Conclusions}\label{sec05}

The creation of a contracting and an expanding pair of universes can be interpreted as the creation of a pair of expanding universes, one filled with matter (the observer's universe) and the other filled with antimatter (the partner universe). It can therefore be seen as the creation of a universe-antiuniverse pair, restoring the apparent matter-antimatter asymmetry observed from the point of view of the single universes. It is worth noticing that the creation of a universe-antiuniverse pair is not necessarily a mechanism for producing the matter-antimatter asymmetry observed in our universe because the particles and antiparticles of the original inflaton field would eventually decay indistinguishable into the particles and antiparticles of the standard model (SM) following the symmetric decays of the SM, so in general we still need a mechanism for producing the baryon asymmetry. However, the creation of universes in universe-antiuniverse pairs does restore the asymmetry because from the global point of view of the two universes the total amount of matter is completely balanced with the amount of antimatter, i.e. whatever is the mechanism producing the baryon asymmetry in one of the universes a parallel mechanism should be producing the antibaryon asymmetry in the partner antiuniverse.

One can also claim that in a multiverse made up of universe-antiuniverse pairs there would be a distribution of universes with different amounts of matter and antimatter, which would be completely balanced however with the amount of antimatter and matter of their partner antiuniverses. In some of these universes the amount of matter and antimatter in each single universe would be balanced too so they would annihilate and these universes would be only full of radiation. These would be perhaps the majority of universes. However, in some universes, due to quantum fluctuations, the amount of matter might slightly exceed the amount of antimatter and those would be the only universes in which galaxies and human being can be produced, being still fully satisfied the matter-antimatter symmetry in the whole multiverse. We would be just living in one of these universes\footnote{I would like to thank M. Dabrowski for suggesting the anthropic version of the matter-antimatter asymmetry produced in a multiverse made up of entangled universe-antiuniverse pairs.}.

Finally, one of the most interesting things of the present proposal is that it provides us with observational imprints of the creation of universes in pairs. The backreaction energy of the matter fields turns out to be observable, at least in principle, but it is not a distinguishable imprint of the existence of a partner antiuniverse. However, the spectrum of fluctuations is modified by the entanglement between the fields of the two universes in such a way that it might produce distinguishable effect, at least in principle, in the observable properties (CMB) of a universe like ours, making testable the creation of universes in universe-antiuniverse pairs and falsifiable the whole multiverse proposal.



\bibliographystyle{apsrev4-1}
\bibliography{../bibliography}

\end{document}